\documentclass[10pt,twocolumn,aps, prb, showpacs,preprintnumbers,amsmath,amssymb]{revtex4-1}

\usepackage{graphicx}
\usepackage{dcolumn}
\usepackage{bm}
\usepackage{color}
\usepackage{soul}
\usepackage{float}

\begin{document}
\newcommand{\Arg}[1]{\mbox{Arg}\left[#1\right]}
\newcommand{\hash}{\mbox{\#\#\#\#\#\#}}
\newcommand{\bb}{\mathbf}
\newcommand{\braopket}[3]{\left \langle #1\right| \hat #2 \left|#3 \right \rangle}
\newcommand{\braket}[2]{\langle #1|#2\rangle}
\newcommand{\br}{\vspace{4mm}}
\newcommand{\bra}[1]{\langle #1|}
\newcommand{\braketbraket}[4]{\langle #1|#2\rangle\langle #3|#4\rangle}
\newcommand{\braop}[2]{\langle #1| \hat #2}
\newcommand{\dd}[1]{ \! \! \!  \mbox{d}#1\ }
\newcommand{\DD}[2]{\frac{\! \! \! \mbox d}{\mbox d #1}#2}
\renewcommand{\det}[1]{\mbox{det}\left(#1\right)}
\newcommand{\ee}{\]} 

\setcitestyle{super}

\newcommand*{\citen}[1]{%
  \begingroup
    \romannumeral-`\x 
    \setcitestyle{numbers}%
    \cite{#1}%
  \endgroup   
}

\preprint{APS/123-QED}

\title{Strain-Modified RKKY Interaction in Carbon Nanotubes}

\author{P. D. Gorman$^{(1)}$}\email{pgorman@tcd.ie}
\author{J. M. Duffy$^{(1)}$}
\author{S. R. Power$^{(2)}$}
\author{M. S. Ferreira$^{(1, 3)}$}

\affiliation{
1) School of Physics, Trinity College Dublin, Dublin 2, Ireland \\
2) Center for Nanostructured Graphene (CNG), DTU Nanotech, Department of Micro- and Nanotechnology, Technical University of Denmark, DK-2800 Kongens Lyngby, Denmark\\
3) CRANN, Trinity College Dublin, Dublin 2, Ireland 
}

\date{\today}

             
\begin{abstract}
For low-dimensional metallic structures, such as nanotubes, the exchange coupling between localized magnetic dopants is predicted to decay slowly with separation.
The long-range character of this interaction plays a significant role in determining the magnetic order of the system.
It has previously been shown that the interaction range depends on the conformation of the magnetic dopants in both graphene and nanotubes.
Here we examine the RKKY interaction in carbon nanotubes in the presence of uniaxial strain for a range of different impurity configurations.
We show that strain is capable of amplifying or attenuating the RKKY interaction,  significantly increasing certain interaction ranges, and acting as a switch: effectively turning on or off the interaction.
We argue that uniaxial strain can be employed to significantly manipulate magnetic interactions in carbon nanotubes, allowing an interplay between mechanical and magnetic properties in future spintronic devices.
We also examine the dimensional relationship between graphene and nanotubes with regards to the decay rate of the RKKY interaction.
\end{abstract}

\pacs{}
                 
\maketitle 
\section{Introduction}
\label{sec:intro}
Recently, there has been much investment in the field of spintronics motivated by the tremendous potential for technological applications. 
Low-dimensional structures such as graphene\cite{Kheirabadi2014123}, nanowires\cite{2011RSPTA.369.3214H}, nanotubes\cite{nt-spin}, nanoribbons\cite{son_half-metallic_2006}, silicene\cite{tsai_gated_2013} (and many more) are expected to lead to useful spintronic applications, possibly leading to the production of extremely efficient magnetic sensors, high-capacity memory storage, and non-volatile computer memories\cite{RevModPhys.76.323,0022-3727-47-19-193001}.
Important to spintronics is the mechanism of interaction between embedded impurities known as the indirect exchange interaction.

\begin{figure}
\includegraphics[width=0.45\textwidth]{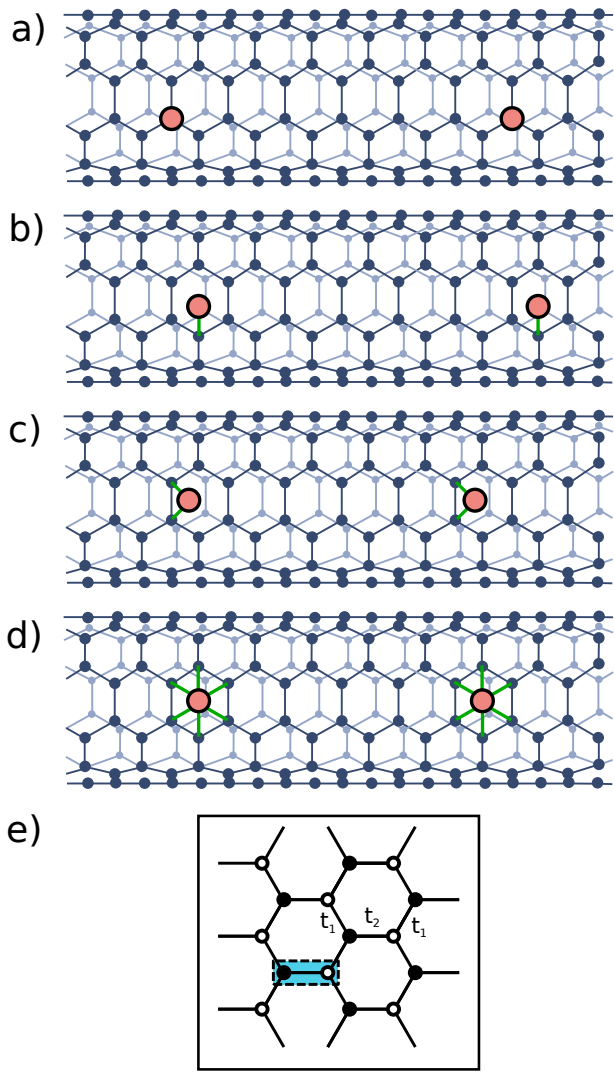}
\caption{
A schematic of an armchair nanotube with impurities, represented by the red circle, whose interactions with the nanotube lattice are indicated by the solid green lines.
Four different types of impurities are represented: a) substitutional impurities, b) top-adsorbed impurities, c) bridge-adsorbed impurities, d) center-adsorbed impurities.
e) A schematic of the unit cell, highlighted in blue, and the inter-atomic hopping integrals $t_2$ and $t_1$.
}
\label{fig:NT_impurities}
\end{figure}

This interaction is one of many such interactions mediated by the conduction electrons of the host material and is realized as the energetically favourable configurations of localized moments, driven by the energy difference between them.
Usually calculated within the Ruderman-Kittel-Kasuya-Yosida (RKKY) approximation~\cite{RKKYIEC,RKKY:RK, RKKY:K,RKKY:Y,RKKY:Bruno1,RKKY:Bruno2}, this interaction has been extensively studied in graphene \cite{stephenreview, Vozmediano:2005, PhysRevLett.97.226801, dugaev:rkkygraphene, saremi:graphenerkky, brey:graphenerkky, hwang:rkkygraphene, bunder:rkkygraphene, rapidcomm:emergence, black:graphenerkky,  uchoa:rkkygraphene, black-schaffer_importance_2010, me:grapheneGF, kogan:rkkygraphene, disorderedRKKY, PhysRevB.84.205409, DynamicRKKY, Peng20123434} and carbon nanotubes~\cite{costa_indirect_2005,kirwan_sudden_2008,klinovaja_rkky_2013,shenoy_rkky_2005,bunder_geometric_2012}, where the focus has been on the sign, magnitude, and rate of decay of the interaction\cite{stephenreview,PhysRevB.90.125411,ourpaper,sherafati:graphenerkky,gorman_rkky_2013, sherafati:rkkygraphene2,duffy_variable_2014}.
This type of analysis has been carried out for the 4 main types of impurities: substitutional, top-adsorbed, bridge-adsorbed, and center-adsorbed (Fig.~\ref{fig:NT_impurities}).
These impurities are differentiated by their conformation with the lattice, which is found to have a strong effect on the behaviour of the RKKY interaction~\cite{black:graphenerkky, sherafati:graphenerkky, uchoa:rkkygraphene, sherafati:rkkygraphene2, kogan:rkkygraphene, disorderedRKKY}.
In carbon nanotubes (CNTs) the RKKY interaction is thought to decay as $D^{-1}$, where $D$ is the separation between impurities.
This is the case for substitutional, top- and bridge-adsorbed impurities, implying that the magnetic moments of adatoms are able to feel their mutual presence even when they are very far apart.
For center-adsorbed impurities (also known as plaquette impurities) the interaction is predicted to have a $D^{-5}$ decay rate~\cite{kirwan_sudden_2008}.
This decay rate is significantly faster than for other impurity types, and is not as useful for applications.
This is unfortunate as it is the preferred configuration for many common Transition Metals, which are likely magnetic dopants\cite{latil_mesoscopic_2004,plaquette111, kengo_nakada_akira_ishii_dft_2011, Foulkes:TB&DFT}.

Controlling the interaction between magnetic objects in carbon nanotubes (CNTs) may provide an entry into the technologically promising area of spintronics.
In order to expand the applicability of those systems we need to understand how the interactions may be modified to suit our needs.
Much progress has been made explaining the intrinsic properties of nanotubes and nearly two decades after their discovery CNTs are still the subject of intensive scientific research due to their intriguing physical properties\cite{ando_electronic_2009,Coleman20061624,SMLL:SMLL201203252}. 
One avenue that has been explored to alter the intrinsic properties of nanotubes is the introduction of strain.
The effects of strain on the elastic, structural, and electronic properties of carbon nanotubes - including band gap and electron conductance - have been extensively studied~\cite{su_electronic_2013,chen_electronic_2008,yang_electronic_2000,sreekala_effect_2008,pullen_structural_2005, ding_curvature_2003,he_effects_2007}.
Many uses have been suggested for strained nanotubes including strain sensors and Field-Effect Transistor~\cite{cullinan_carbon_2010,yoon_analysis_2007}.
It has been shown that uniaxial strain can open a bandgap in metallic nanotubes and alter the bandgap in semi-conducting nanotubes~\cite{minot_tuning_2003,fang_strain-induced_2011}.
One avenue that has not been explored is how strain may affect their magnetic properties.

It has been shown that the properties of RKKY interaction in both graphene \cite{uchoa:rkkygraphene,ourpaper,gorman_rkky_2013} and nanotubes\cite{costa_indirect_2005,kirwan_sudden_2008} are strongly influenced by the conformation of the magnetic impurities with the lattice.
In graphene it has been shown that uniaxial strain has the ability to amplify or attenuate that interaction, as well massively increase the range or suppress the interaction in some cases~\cite{ourpaper,PhysRevB.87.155431,gorman_rkky_2013,PhysRevB.80.245436}.
In graphene nanoribbons strain has been shown to tune the exchange splitting of non-vanishing moments induced by vacancies in the lattice\cite{midtvedt_strain-tuning_2015}.
In this paper we determine the role that strain plays in controlling the RKKY interaction in achiral carbon nanotubes, and explore the possibility of using strain to modify the interaction between such impurities.
We examine the four main types of impurity conformations: substitutional, where a carbon atom is replaced by an impurity; top-adsorbed, where an impurity sits above, and connects to, a single carbon atom; bridge-adsorbed, where an impurity sits above the bond between two neighbouring carbon atoms; and center-adsorbed, where an impurity sits at the center of a carbon hexagon and attaches equally to all six surrounding carbons.
This work is carried out for both armchair-edged nanotubes (ACNT) and achiral zigzag-edged nanotubes (ZZNT) as they have substantially different reactions to strain.
Figure~\ref{fig:NT_impurities} schematically shows the four conformation types in an achiral ACNT, where the impurities are separated along the axial direction of the tube.
We provide simple expressions for the interactions between two impurities on a strained nanotube. 
The simplicity of these expressions allow us to understand the modified coupling in terms of the inter-carbon hopping integrals, and analytically understand how strain may change the decay rate drastically.
We also explore the role played by nanotube circumference in determining the decay rate of the RKKY interaction between impurities in order to validate our model.

\section{Methods}
\label{sec:meth}

The energy difference between the ferromagnetic (FM) and antiferromagnetic (AFM) alignments of two moments embedded in a conducting host is described as the indirect exchange coupling between the two moments.
The Lloyd formula method allows us to calculate the total energy difference, $J_{AB}$, between two magnetic impurities labelled $A$ and $B$.
The Lloyd formula is given by
\begin{equation}
J_{AB} = -\frac{1}{\pi} \int dE \ f(E) \ln{ 
\left({1+ 4 {V_{ex}^2} \ {g}_{ab}(E){g}_{ba}(E)}\right)
} 
\label{eq:Lloyd} 
\end{equation}
where $f(E)$ is the Fermi function, $V_{ex}$ is a spin-dependent onsite potential that accounts for the exchange splitting in the magnetic orbitals, and $\mathcal{G}_{AB}^{\sigma} (E)$ is the real-space, single-electron Green Function (GF) describing the propagation of electrons with spin $\sigma = \uparrow$ or $\downarrow$.
According to our definition $J<0$ ($J>0$) corresponds to a ferromagnetic (anti-ferromagnetic) alignment of magnetic moments.
Our nanotube is modelled using the Nearest Neighbour Tight Binding Approximation with an inter-atomic hopping integral $t_0= -2.7 eV$, which provides a good approximation of the electronic structure of CNTs for all but the smallest circumference tubes.
Unless otherwise specified we will use $t_0$ as our energy unit.

We assume that each substitutional impurity causes a change in onsite energy, and that each adsorbed impurity orbital has a finite hopping $\tau$ to $N$ adjacent sites on the graphene lattice: top-adsorbed ($N=1$), bridge-adsorbed ($N=2$), and center-adsorbed ($N=6$).
We will only consider a single magnetic orbital at each impurity site, separated along the longitudinal direction of the nanotube, however, it is straightforward to generalize this approach to deal with multiple orbitals, or separations with an axial component.
We will not deal with the exact parametrizations for specific impurity types here.
These parametrizations can be found in numerous \emph{ab initio} studies\cite{latil_mesoscopic_2004, kengo_nakada_akira_ishii_dft_2011, Foulkes:TB&DFT} of single impurities in nanotubes with different configurations, though it is worth noting that the results presented here are not strongly dependent on the impurity type used.

Strain is introduced into the nanotube system, in a similar manner as for a graphene system\cite{pereira_tight-binding_2008,pereira_strain_2009,ourpaper,gorman_rkky_2013}, by splitting the hopping integral between carbon atoms, $t_0$, into $t_2$ and $t_1$.
In the convention used here $t_2$ is the intra-unit-cell hopping and $t_1$ is the inter-unit-cell hopping (Fig.~\ref{fig:NT_impurities}e).
The strain-dependent, real-space GF between two sites on the graphene lattice can be written as a double integral over the Brillouin zone\cite{costa_indirect_2005,me:grapheneGF} in the form
\begin{equation}
 g_{ab} = \frac{1}{2\pi^2} \int\limits_{-\pi/2}^{\pi/2} dk_Z \int\limits_{-\pi}^{\pi} dk_A \, \frac{N_{\gamma}(E,\mathbf{k}) \, e^{i \mathbf{k} \cdot \mathbf{D}}}{E^2 - \, |f(\mathbf{k})|^2},\\
  \label{eq:gab_int}
\end{equation}
where
\begin{equation}
  f(\mathbf{k}) = t_2 + 2 t_1 \cos(k_Z)e^{ik_A}.
\end{equation}
Here $N_{\gamma}$ contains information about the sites of each impurity, $D$ is a separation vector, and $f(\bf{k})$ is a sum of Bloch phase terms over nearest neighbours.

The Green's function (GF) for nanotubes is derived in a similar manner to that of the GF for pristine graphene, except that the periodicity in the nanotube causes one of the components of one of the k-vectors to be quantized.
The integral then becomes a sum over all the unique k-points, which are obtained from the quantization condition which describes the circumference of the nanotube, $n_c$.
Following the procedure of Ref.~\citen{costa_indirect_2005} the GFs for both ACNTs and ZZNTs may now be written in a general form
\begin{equation}
 \mathcal{G}_{AB} = \sum_j \mathcal{A}(E,j) e^{i \mathcal{Q}(E,j)D},
\end{equation}
which will allow us to solve the Lloyd formula analytically.

To calculate the coupling between two impurities, $A$ and $B$, we now use the RKKY approximation to write Eq. \ref{eq:Lloyd} as
\begin{equation}
 J \sim \text{Im} \int dE f(E) V_{ex}^2 \mathcal{G}_{AB}^2.
\end{equation}
Using the general form of our GFs we can simplify our coupling to
\begin{equation}
 J \sim \sum_{j,k} \text{Im} \int dE f(E) V_{ex}^2  \mathcal{B}(E,j,k) e^{i 2\mathcal{Q}(E,j,k)D},
\end{equation}
where $\mathcal{B}(E,j,k) = \mathcal{A}(E,j) \mathcal{A}(E,k)$.

We have shown previously that the magnetic coupling can be easily extracted when the coupling is expressed in such a form, by reducing the integration to a sum over Matsubara frequencies and expanding the functions $\mathcal{B}(E)$ and $\mathcal{Q}(E)$ around the Fermi energy in the low temperature limit\cite{ourpaper,gorman_rkky_2013}.
Note that $\mathcal{B}(E)$ and $\mathcal{Q}(E)$ are also strain-dependent, but the variable, $\varepsilon$, has been omitted for neatness.
The coupling between impurities can then be expressed as
\begin{equation}
\label{eq:J-final}
 J \sim  - V_{ex}^2 \sum_{j,k} \text{Im}
 \left({
  e^{2 i \mathcal{Q}^{(0)}D} 
  \sum_l
  \left[{
    \frac{\mathcal{B}^{(l)}}{(2i\mathcal{Q}^{(1)})^{l+1}}
    \frac{1}{D^{(l+1)}}
  }\right]
 }\right),
\end{equation}
where the dependence of $\mathcal{Q}$ and $\mathcal{B}$ on $E,j,k$ have been omitted for simplicity.


\section{RKKY Interaction as a function of nanotube circumference}

It has previously been reported that in nanotubes the RKKY interaction between substitutional, top-adsorbed, and bridge-adsorbed impurities decays as $D^{-\alpha}$ with $\alpha =1$, and $\alpha = 5$ for center-adsorbed impurities, where $D$ is the separation between the impurities\cite{kirwan_sudden_2008}.
However these have been calculated without much regard for the effect that the circumference of the nanotube has on the interaction.
In bulk graphene these rates are found to be $\alpha=3$ for substitutional, top-adsorbed, and bridge-adsorbed, and $\alpha=7$ for center-adsorbed\cite{gorman_rkky_2013}.
In this section we examine the role of the nanotube circumference, $n_c$ (which can be easily related to the more usual measure of diameter), for both armchair nanotubes and metallic zigzag nanotubes, to understand how the RKKY interaction transitions from the behaviour predicted in nanotubes to that predicted in bulk graphene.
Here $n_c$ is a dimensionless quantity which describes the number of vectors between unit cells, in the armchair (zigzag) direction for ACNTs (ZZNTs), that are required to traverse the circumference of the nanotube.

\begin{figure}[h]
\includegraphics[width=0.45\textwidth]{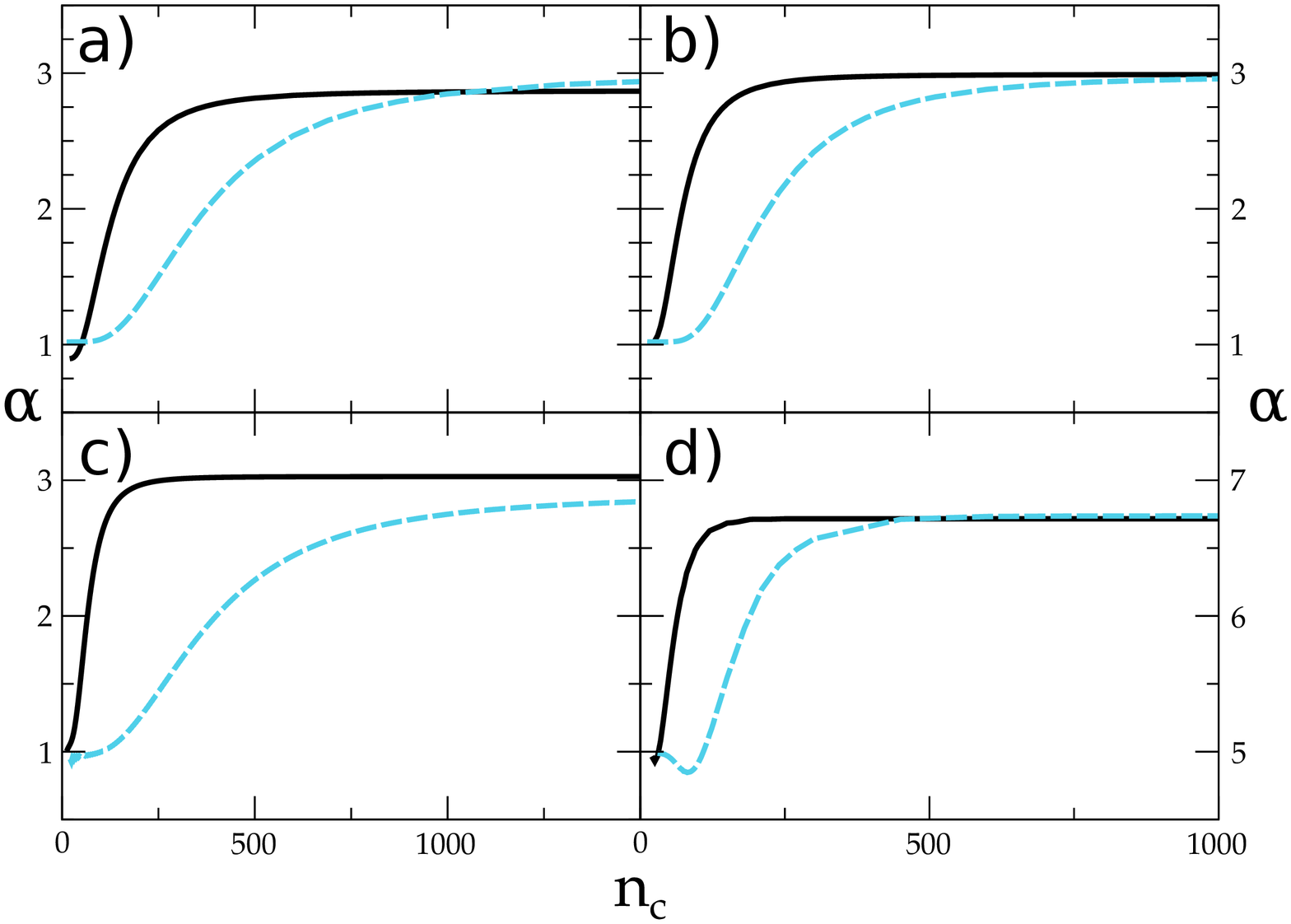}
\caption{
Decay ($\alpha$) versus circumference ($n_c$) for ACNT (solid-black) and ZZNT (dashed-blue) for
a) substitutional impurities, same-sublattice,
b) substitutional impurities, opposite-sublattice,
c) bridge-adsorbed,
d) center-adsorbed,
}
\label{fig:NT_decay_vs_circumference_ALL}
\end{figure}

Figure~\ref{fig:NT_decay_vs_circumference_ALL} shows the change in decay, $\alpha$, between impurities as a function of the circumference for both ACNT and ZZNT at $E_F=0$.
For  both same- and opposite- lattice substitutional impurities, and bridge-adsorbed impurities we see that the greatest predictor of behaviour is whether the nanotube is an ACNT or a ZZNT.
On ACNTs the interaction very quickly decays away from the nanotube case of $\alpha=1$ to the graphene case of $\alpha=3$.
This is seen in \ref{fig:NT_decay_vs_circumference_ALL} a), b), c) as the black line, which resembles graphene by about $n_c = 200$.
On ZZNTs the interaction has a plateau of NT behaviour before moving gradually to the behaviour seen in graphene, this is shown as the blue line, and does not resemble graphene until about $n_c = 1000$.

Figure \ref{fig:NT_decay_vs_circumference_ALL} d) shows the same effect for center-adsorbed impurities.
By increasing the circumference of the NT we show that both ACNT and ZZNT decay rates go from $D^{-5}$ to $D^{-7}$. 
Again, impurities on ACNT decay much more rapidly towards the bulk case and on ZZNT exhibit a plateau and a slow decay towards bulk.
In ACNTs the axial separation is along the zigzag direction, which is known to feature a repeating triplet pattern of interaction strength: strong, strong, weak\cite{costa_indirect_2005}.
The circumference was increased in steps of three to produce a smooth curve.
This result shows a geometry dependence in the transition from nanotube-like to graphene-like behaviour as the diameter of the CNT increases, and may be relevant for studies of the largest feasible CNTs since there is a notable difference between the two geometries even at quite small $n_c$.

\section{Axial strain effects on RKKY interaction in nanotubes}

\begin{figure}
\includegraphics[width=0.45\textwidth]{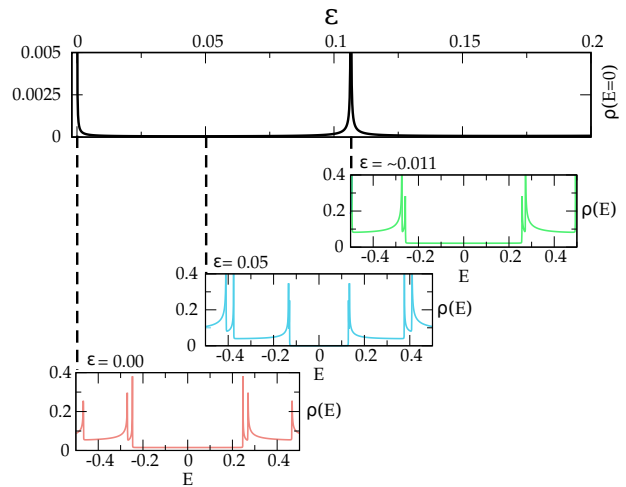}
\caption{
The density of states, $\rho$, at E = 0 for a ZZNT with $n_c=21$ under uniaxial strain, $\varepsilon$.
Subfigures show the density of states against energy for three strain values: $\varepsilon = 0$ (red), $\varepsilon=0.05$ (blue), and $\varepsilon \approx 0.11$ (green).
}
\label{fig:ZZNT_DOS_with_strain}
\end{figure}

Unstrained achiral nanotubes may be metallic or semiconducting.
ACNTs are known to be metallic for all circumferences, $n_c$, whereas ZZNTs are only metallic if $n_c=3k$, where $k$ is an integer, and are otherwise semi-conducting.
These properties are due to the intersection of the discretized momentum values and the Fermi surface.
The introduction of strain, $\varepsilon$, into the nanotube system alters the coincidence of the momentum values and the Fermi surface, this in turn changes the electronic properties of nanotubes.
ACNTs remain metallic under strain, however strain can change a conducting ZZNT into a insulating one, and conversely an insulating one into a conducting one\cite{minot_tuning_2003}.
Figure \ref{fig:ZZNT_DOS_with_strain} shows the density of states (DOS) of a ZZNT with $n_c=21$ as strain is applied.
The initially metallic NT immediately becomes semi-conducting with slight strain, and as strain is further applied the two peaks in density of states come together and eventually merge.
This is seen in all circumferences of semi-conducting ZZNTs, where the precise value of strain required to shift the density of states away from zero around $E_F=0$ decreases as the circumference increases.

\subsection{Armchair-edged Nanotubes}
\begin{figure}
\includegraphics[width=0.45\textwidth]{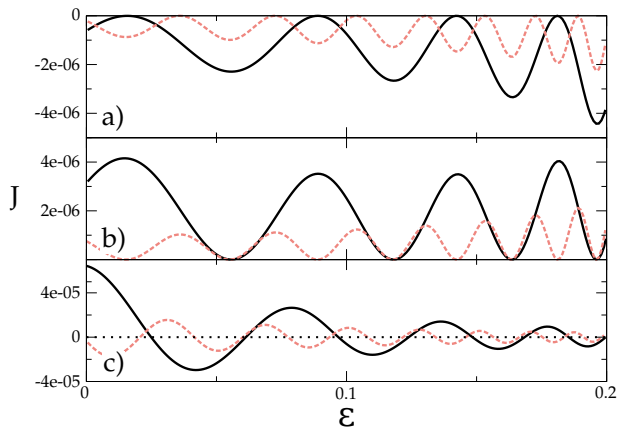}
\caption{
Plot of the coupling (J) as a function of strain ($\varepsilon$) for ACNTs at D = 10 (black) and D = 20 (dashed-red) for a) same-sublattice substitutional, b) opposite-sublattice substitutional, and c) bridge-adsorbed impurities, where the dotted-black line indicates the transition from FM to AFM.
}
\label{fig:ACNT_rkky_vs_strain_sub_bridge}
\end{figure}

Although ACNTs remain metallic with strain, this does not mean that the RKKY interaction is unchanged.
Figure \ref{fig:ACNT_rkky_vs_strain_sub_bridge} shows the effect of strain on the RKKY interaction for substitutional and bridge-adsorbed impurities for several separations $D$, on an ACNT with $n_c = 21$.
Top-adsorbed impurities have been omitted due to the similarity of behaviour with substitutional impurities.
In all three cases oscillations are seen, but only in c) are these oscillations around zero signifying a change from ferromagnetic to anti-ferromagnetic behaviour.
For both a) same-sublattice and b) opposite-sublattice cases, strain is seen to either amplify or attenuate the interaction depending on the precise value of strain.
This allows for the magnitude of these interactions to be precisely controlled.
Interestingly there exist strain values (well within the experimental limits) that completely shut off the opposite-sublattice interaction, whilst simultaneously maximizing the same-sublattice interaction.
In addition to this strain seems to generally amplify the interaction, while leaving the decay rate unchanged.
This is understood from Eq.~\ref{eq:J-final}.
The overall decay rate can be found from the first non-vanishing $\mathcal{B}(\varepsilon)$, which, for same-sublattice substitutional impurities, is
\begin{equation}
 {\mathcal{B}(\varepsilon)}^{(0)} = \frac{1}{4{n_c}^2( t_2^2 -4t_1^2 )}.
\end{equation}
This factor corresponds to a coupling decay rate of $\alpha={1}$, identical to that of the unstrained case.
The equivalent opposite-sublattice term is similar.
Small values of strain will not cause this term to vanish, and so no change in the decay rate is expected.

For bridge-adsorbed impurities (Fig.~\ref{fig:ACNT_rkky_vs_strain_sub_bridge} c) strain is capable of switching the sign of an interaction in a very controlled fashion.
Since bridge-adsorbed impurities display a coupling with alternating sign as a function of separation strain cannot create wholly FM or AFM interactions between multiple impurities.
However, for a pair of impurities it is capable of switching the sign of the interaction.
The coupling between bridge-adsorbed impurities is weakened with strain, however the decay rate is not affected.
The first non-vanishing $\mathcal{B}(\varepsilon)$ is
\begin{equation}
 {\mathcal{B}(\varepsilon)}^{(0)} = \frac{4}{{n_c}^2( t_2^2 -4t_1^2 )},
\end{equation}
again corresponding to a decay rate of $D^{-1}$ for small strains.

\begin{figure}
\includegraphics[width=0.45\textwidth]{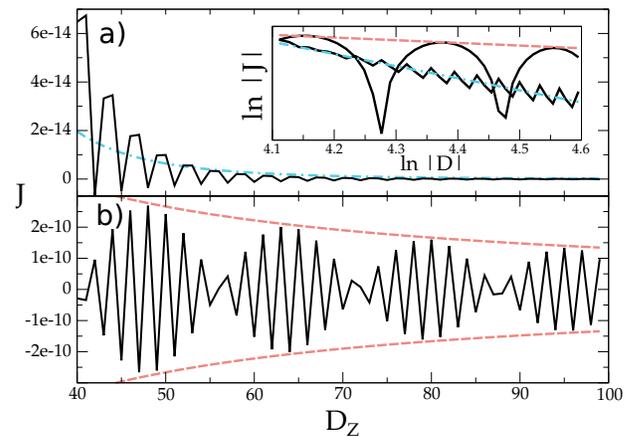}
\caption{
The interaction between two centre-adsorbed impurities in an ACNT in a) the absence of strain ($\varepsilon = 0$), and b) the presence of strain ($\varepsilon = 0.1$).
The inset is a normalized log-log plot of both interactions.
The coloured lines show the decay rates: $D^{-5}$ (dash-dotted-blue), $D^{-1}$ (dashed-red).
}
\label{fig:ACNT_rkky_vs_strain_center}
\end{figure}

For center-adsorbed impurities we see that the introduction of strain has a large impact on the decay rate and hence the magnitude of the interaction.
Figure \ref{fig:ACNT_rkky_vs_strain_center} shows the behaviour of the interaction between center-adsorbed impurities with and without strain.
Strain changes the interaction from being primarily antiferromagnetic to being split evenly between ferromagnetic and antiferromagnetic.
The interaction between these impurities is known to decay rather quickly as $D^{-5}$ (Fig.~\ref{fig:ACNT_rkky_vs_strain_center} a).
This can be understood from the ${\mathcal{B}(\varepsilon)}^{(l)}$ terms, which are zero for $l = 0, 1, 2, 3$.
The first non-zero term is found at ${\mathcal{B}(\varepsilon)}^{(4)}$, (it is too long to reproduce here) and thus indicates the known decay rate of $D^{-5}$.
The small values here owe to the necessity of examining the coupling at large ranges to ascertain a decay rate.

However, the introduction of strain breaks a key symmetry of the system massively increasing the range of the interaction.
The interaction in the presence of strain decays slowly as $D^{-1}$ (Fig.~\ref{fig:ACNT_rkky_vs_strain_center} b).
The inset of figure \ref{fig:ACNT_rkky_vs_strain_center} shows a log-log plot of the coupling between the impurities for the unstrained and strained cases.
Examining the ${\mathcal{B}(\varepsilon)}^{(0)}$ we find an explanation for the increased range,
\begin{equation}
 {\mathcal{B}(\varepsilon)}^{(0)} = \frac{(t_2-t_1)^4 \phi(\varepsilon)}{4{n_c}^2t_1^4(t_2^2 - 4t_1^2)},
\end{equation}
where $\phi(\varepsilon)$ is a strain-dependent phase term.
This simple expression allows us to understand the role that strain plays on the decay rate of centre-adsorbed impurities
Here the numerator displays a $t_2-t_1$ term, since the unstrained system is defined by $t_2=t_1$ it is exactly zero only in unstrained graphene.
Such a non-zero ${\mathcal{B}(\varepsilon)}^{(0)}$ term corresponds to a decay rate of $D^{-1}$.
This symmetry breaking effect is analogous to that seen in graphene where the decay rate is massively reduced from $D^{-7}$ to $D^{-3}$ by the introduction of strain\cite{gorman_rkky_2013}.
This effect allows the strength of the interaction to be massively increased with a relatively small amount of strain.

\subsection{Zigzag-edged Nanotubes}
\begin{figure}
\includegraphics[width=0.45\textwidth]{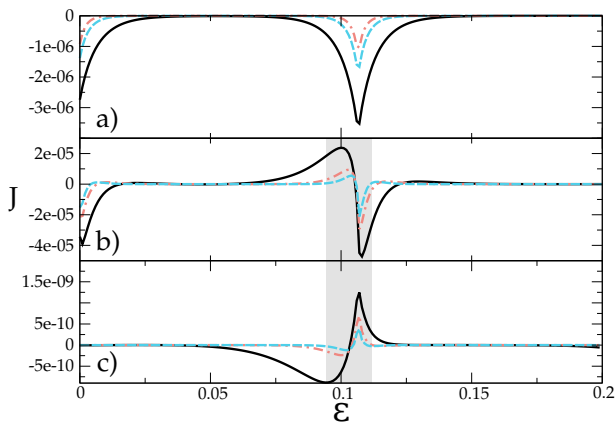}
\caption{
The coupling between a) substitutional, b) bridge-adsorbed, and c) centre-adsorbed impurities is plotted as a function of strain in a ZZNT with $n_c=21$.
Each coupling is plotted for three values of separation $D=10$ (solid-black), $D=20$ (dashed-red), $D=30$ (dash-dotted-blue).
The grey bands indicate the regions of sudden FM/AFM switch.
}
\label{fig:ZZNT_rkky_versus_strain_all}
\end{figure}

We now move our attention to ZZNTs.
Since the introduction of strain changes a conducting ZZNT to a semi-conducting ZZNT the effect of strain in ZZNTs is markedly different.
Figure \ref{fig:ZZNT_rkky_versus_strain_all} shows the coupling between substitutional, bridge-adsorbed, and center-adsorbed impurities in ZZNT with $n_c=21$ as a function of strain.
Since the introduction of strain changes a ZZNT from conducting to semi-conducting the RKKY interaction in an initially conducting ZZNT goes to zero (Fig.~\ref{fig:ZZNT_rkky_versus_strain_all} a).
Strain also has the opposite effect on initially semi-conducting ZZNTs, taking the RKKY away from zero for precise values of strain.
Similar behaviour is seen for bridge-adsorbed impurities (Fig.~\ref{fig:ZZNT_rkky_versus_strain_all} b), with the key difference being the sign change that occurs around this precise strain value.
What this means is that for two bridge-adsorbed impurities separated by some distance $D$ there exists a small range of strain that can quickly turn the RKKY interaction from strongly AFM to strongly FM.
This is in contrast to ACNTs where a sign change can be achieved but only gradually.
The precise strain value will depend on the circumference of the nanotube $n_c$.
In addition, the sign change occurs at all separations, meaning that it is possible for strain to quickly change a system of $N$ impurities from all having the same alignment to having alternating alignments leading to no overall magnetic moment. 
The center-absorbed impurity (Fig.~\ref{fig:ZZNT_rkky_versus_strain_all} c) shows similar behaviour, however since a key symmetry that leads to the $D^{-5}$ decay is broken by strain the magnitude is increased massively, which gives the appearance of initially being zero.
This could prove extremely useful for spintronic applications where it allows minor changes in strain to result in a large change in the magnetic ordering of impurities.
This change could be implemented in a reversible and controllable manner.

\section{Conclusions}
\label{sec:conc}
In summary, we have presented simple expressions for the RKKY interaction between magnetic impurities in strained nanotubes.
We first validated our model by examining the role that the circumference of the nanotube plays in the decay rate as the nanotubes become more graphene-like, and found that ZZNT maintain their nanotube-like behaviour for larger circumferences than ACNT.
We have also shown that strain can change the magnitude, sign, and decay of these interactions in ACNTs.
The amplification was particularly pronounced for centre-adsorbed impurities  where we showed that the symmetry breaking of the hexagonal lattice by uniaxial strain leads to a significantly slower decay rate between centre-adsorbed impurities: $D^{-5}$ to $D^{-1}$ in armchair nanotubes.
Zigzag nanotubes, meanwhile, display a wide range of amplification and switching effects with minor variations of the applied strain. 
These features are related to the transition from metallic to semiconducting behaviour.
Experiments to date searching for magnetism in disordered graphene seem to suggest paramagnetic, non-interacting moments~\cite{sepioni_limits_2010}. 
Signatures of indirect exchange interactions between such moments in graphene are very difficult to detect due to their short-ranged nature, particularly if they adopt certain adsorption configurations.
Amplification of these couplings using strain may provide a path to their detection in future experiments.
The abrupt changes in interaction strength and behaviour, demonstrated here for ZZNTs, suggest that strained nanotubes can act as switches for magnetoresistive behaviour, which may find use in future spintronic devices.
Thus strain presents itself as a way to control interactions between magnetic dopants in nanotubes, and the overall magnetic ordering of the material, in a reversible and controllable manner - an indispensable trait for future spintronic devices.


\begin{acknowledgments}
The authors acknowledge financial support received from the Programme for Research in Third-Level Institutions PRTLI5 Ireland, the Irish Research Council for Science, Engineering and Technology under the EMBARK initiative and from Science Foundation Ireland under Grant No. SFI 11/RFP.1/MTR/3083. 
The Center for Nanostructured Graphene (CNG) is sponsored by the Danish National Research Foundation, Project No. DNRF58.
\end{acknowledgments}

\end{document}